\documentclass[twocolumn,secnumarabic,amssymb, nobibnotes, aps, prd]{revtex4}
\usepackage{graphicx}

\begin{document}
\title{Bohm's quantum potential and quantum force in superconductor\footnote{Invited talk presented at the conference "Foundations of Probability and Physics-5" Vaxjo University, Sweden, August 24-27, 2008.}}
\author{A.V. Nikulov}
\affiliation{Institute of Microelectronics Technology and High Purity Materials, Russian Academy of Sciences, 142432 Chernogolovka, Moscow District, RUSSIA.} 
\begin{abstract} The Bohm's quantum potential, introduced in 1952, and  the quantum force in superconductor, introduced in 2001, allow to describe non-local force-free momentum transfer observed in the Ahronov-Bohm effects. Comparison of the Ahronov-Bohm effects in the two-slit interference experiment and in superconductor ring reveals fundamental difference between the Schrodinger wave function and the wave function describing macroscopic quantum phenomena. The Ginzburg-Landau wave function describing the superconductivity phenomenon can not collapse and an additional postulate, which was implied first by L.D. Landau, must be used for the description of macroscopic quantum phenomena. It is note that quantum principles and postulates should not be universal till the quantum formalism is only phenomenological theory but no description of an unique reality. A simple Gedankenexperiment is considered which challenges the universality of the Heisenberg uncertainty relation and the Bohr's complementarity principle.
 \end{abstract}

\maketitle

\narrowtext

\section*{Introduction}

Quantum mechanics, in contrast to other theories of physics, has interpretations. Moreover it has many interpretations. The quantum mechanics, or to be more precise quantum formalism, describes, in contrast to other physical theories, no unique reality but only phenomena. By definition, a phenomenon (from Greek $\varphi \alpha \iota \nu \omega \mu \epsilon \nu \omega \nu $, pl. $\varphi \alpha \iota \nu \omega \mu \epsilon \nu \alpha  $ - phenomena) is any occurrence that is observable. In other words a phenomenon is that what we can observe. For centuries scientists believed that phenomena reveal elements of an objective reality and the aim of the theory is a description of this objective reality. The theory describing the unique objective reality can not have interpretations. It can describe the reality correctly or not correctly, and there can be no interpretation whatever. But quantum phenomena turned out so paradoxical that no universally recognized interpretation of they as manifestation of an unique objective reality can be created up to now. The quantum formalism created by Planck, Einstein, Bohr, Heisenberg, de Broglie, Schrodinger, Born and others has been enormously successful in describing numerous quantum phenomena. But there is no generally accepted answer on the question: "What could these quantum phenomena manifest?"  

The absence of such answer provoked discussion between Albert Einstein \cite{EPR35,Einste49,Einstein} and Niels Bohr \cite{Bohr35,Bohr49,Bohr}, which is in progress among expersts up to now, in particular at the Vaxjo Conferences \cite{Vaxjo}. These discussions and  the plurality of interpretations point out to our incomprehension of subject of the universally recognized quantum description. In this work I would like first of all to expand the region of this incomprehension. The quantum formalism was developed for description of atomic phenomena and the controversy about its interpretation concerned for the present exclusively this level \cite{EPR35,Einstein,Bohr35,Bohr,Vaxjo}. The macroscopic quantum phenomena, such as superconductivity and superfluidity, were not touched upon almost. There can be no doubt that superconductivity and superfluidity are quantum phenomena. But could we be sure that the principles of quantum formalism challenging to local realism, such as superposition and wave function collapse, can be applied to these phenomena? Could we be sure that the Ginzburg-Landau wave function describing  superconducting state has the same nature as the Schrodinger wave function? In order to answer on these questions I will compare the Bohm's quantum potential, introduced in 1952 \cite{Bohm1952}, and  the quantum force in superconductor, introduced in 2001 \cite{QF2001}, and will consider the Ahronov-Bohm effects \cite{Bohm1959} in the two-slit interference experiment and in superconductor ring. The non-universality revealed in fundamental difference between the Schrodinger and Ginzburg-Landau wave functions may be general feature of the quantum formalism as phenomenological theory. 

\section{Schrodinger wave function and Ginzburg-Landau wave function}
The Schrodinger wave function $\Psi _{Sh} =|\Psi _{Sh}|\exp{i\varphi }$ was introduced 1926 for description of atomic phenomena and the Ginzburg-Landau one $\Psi _{GL} =|\Psi _{GL}|\exp{i\varphi }$ was introduced 1950 \cite{GL1950} for description of superconductivity phenomena. The both theories are phenomenological as well as the whole quantum formalism. The both wave functions have allowed to describe the discrete spectrum, the first one of atom and the second one of superconductor ring. The phase $\varphi $ of the wave functions has the same interpretation in the both theories: 
$$\hbar \nabla \varphi = p = mv + qA \eqno{(1)}$$ 
is the canonical momentum of a particle with a mass $m$ and a charge $q$ in the presence of a magnetic vector potential $A$. But in contrast to the Born's interpretation the $|\Psi _{GL}|^{2}$ value is no a probability but a density $|\Psi _{GL}|^{2} = n_{s}$. When the density is constant in the space $\nabla n_{s} = 0$ the superconducting current density   \cite{Tinkham}
$$j = \frac{q}{m}|\Psi _{GL}|^{2}(\hbar \nabla \varphi - qA) \eqno{(2)}$$ 
may be written as the product $j = qn_{s}v$ of the density $n_{s}$ of particles with a charge $q$ and their velocity $v = (\hbar \nabla \varphi - qA)/m$. L.D. Landau wrote the relation (2) as far back as 1941, in the work \cite{Landau41} explaining superfluidity phenomena. 

\subsection{Macroscopic quantum phenomena}
The relation (2) together with the GL wave function allows to describe the majority of the macroscopic quantum phenomena observed in superconductors. Since the complex wave function $\Psi _{GL} =|\Psi _{GL}|\exp{i\varphi }$ must be single-valued at any point in the superconductor, its phase $\varphi $ must change by integral multiples $2\pi$ following a complete turn along the path $l$ of integration and consequently  
$$\oint_{l}dl \nabla \varphi = \oint_{l}dl p/\hbar  = n2\pi \eqno{(3)}$$ 
The relation (3) is the quantization of the angular momentum postulated by Bohr as far back as 1913. According to the relations (2), (3) and  $\oint_{l}dl A = \Phi $ the integral of the current density along any closed path inside superconductor 
$$\mu _{0}\oint_{l}dl \lambda _{L}^{2} j  + \Phi = n\Phi_{0}  \eqno{(4)}$$  
must be connected with the integral quantum number $n$ and the magnetic flux $\Phi $ inside the closed path $l$. $\lambda _{L} = (m/\mu _{0}q^{2}n_{s})^{0.5} = \lambda _{L}(0)(1 - T/T_{c})^{-1/2}$  is the quantity generally referred to as the London penetration depth \cite{London35}; $\lambda _{L}(0) \approx 50 \ nm = 5 \ 10^{-8} \ m$ for most superconductors \cite{Tinkham}; $\Phi _{0} = 2\pi \hbar /q$ is the quantity called flux quantum. According to (4) and the Maxwell equation $curl H = j$ the current density decreases strongly inside superconductor 
$$j = j_{0}\exp{-\frac{|r - r_{1}|}{\lambda _{L}}} \eqno{(5)}$$  
where the coordinate $r$ run from the surface (at $r = r_{1}$) into the interior $r < r_{1}$ of the superconductor with the radius $r_{1}$. 

\subsubsection{Magnetic flux quantization and the Meissner effect}
The flux quantization $\Phi = n\Phi _{0}$ discovered experimentally in 1961 \cite{FQ1961} is observed since $j = 0$ inside superconductor $ r_{1} - r \gg \lambda _{L}$ according to (5). The Meissner effect i.e. the expulsion of magnetic flux $\Phi $ from the interior of a superconductor, discovered by Meissner and Ochsenfeld in 1933 \cite{Meissner}, may be considered as a particular case $n = 0$ of the flux quantization $\Phi = n\Phi _{0} = 0$. The quantum number $n$ must be equal zero when the wave function $\Psi _{GL} =|\Psi _{GL}|\exp{i\varphi }$  has no singularity inside $l$, since the radius $r$ of the integration path $l = 2\pi r$ can be decreased down to zero in this case. The Meissner effect is first macroscopic quantum phenomenon observed experimentally. Much more marvellous one is observed in the Abrikosov state discovered experimentally as far back as 1935 \cite{MixSt} and described theoretically by A.A. Abrikosov in 1957 \cite{Abrikos}. The magnetic flux can penetrate the interior of  type-II superconductor with the Abrikosov vortex \cite{Huebener} which is the singularity of the GL wave function corresponding to $n = 1$ in (3). Numerous direct observations \cite{Huebener,VortexEx} give evidence that $\Phi = n\Phi _{0}$ in any macroscopic region, where $\Phi $ and $n$ are the magnetic flux and the number of the Abrikosov vortices observed inside $l$. 
 
\subsubsection{The velocity quantization and the persistent current}
The flux quantization \cite{FQ1961} and the Meissner effect \cite{Meissner} are observed at strong screening when the wall thickness $w = r_{1} - r_{2} \gg  \lambda _{L}$, where $r_{1}$ and $r_{2}$ are the outside and inside radius of cylinder or ring. At weak screening $w \ll  \lambda _{L}$ the quantization of the persistent current \cite{PerCur61} $I_{p} = sj = sqn_{s}v$ (4) or the velocity $v$ of superconducting pairs 
$$\oint_{l}dl v  =  \frac{2\pi \hbar }{m}(n - \frac{\Phi}{\Phi_{0}})  \eqno{(6)}$$
is observed. $v = (\hbar /rm)(n - \Phi /\Phi _{0})$ in a ring with radius $r$ and uniform section $s$. The persistent current is a periodical function $I_{p}(\Phi /\Phi _{0})$ of the magnetic flux $\Phi$ inside the ring with period equal the flux quantum $\Phi _{0}$ under equilibrium conditions corresponding to minimum energy, i.e. the minimum of the $v^{2} \propto (n - \Phi /\Phi _{0})^{2}$. The quantum oscillations of the persistent current $I_{p}(\Phi /\Phi _{0})$ are observed not only in superconducting state but also at non-zero resistance $R > 0$, in the fluctuation region of superconductor ring \cite{PC1997} and also in normal metal \cite{normal} and semiconductor \cite{semi,semi10nm} mesoscopic rings. The Little-Parks effect \cite{LP1962} is first experimental evidence of the persistent current observed at $R > 0$. 

\subsection{Quantum force}
The numerous observations \cite{PC1997,normal,semi,semi10nm,LP1962} of the $I_{p}(\Phi /\Phi _{0})$ at $R > 0$ agree with theory \cite{Kulik70,PCteor} obtained in the limits of the universally recognized quantum formalism. But the observations of the direct circular current $I_{p} \neq 0$ at $R > 0$ and $d\Phi /dt = 0$ display obvious absent of the force balance. The conventional direct circular current $I$ can be observed a long time at $R > 0$ only at a non-zero Faraday's voltage $RI = -d\Phi /dt$, when the force $qE$ of the electric field $E = -dA/dt$ counterbalances the dissipation force. This current in a ring with an inductance $L$ and a resistance $R > 0$ must disappear during the relaxation time $\tau _{RL} = L/R $ at $d\Phi /dt = 0$. But the persistent current does not disappear. Moreover the permanent potential difference $V_{dc}(\Phi /\Phi _{0}) \propto I_{p}(\Phi /\Phi _{0})$ is observed on a system of asymmetric rings \cite{Letter07}. It is observed in \cite{Letter07} that the persistent current $I_{p}(\Phi /\Phi _{0})$ flows against the electric field $E = -\nabla V_{dc}(\Phi /\Phi _{0})$ in one of the semi-rings because of the absence of the Faraday's circular electric field $E = -\nabla V - dA/dt = -\nabla V$ at $d\Phi /dt = 0$. The potential electric field $E = -\nabla V_{dc}(\Phi /\Phi _{0})$ is directed, for example, from left to right at $\Phi = \Phi _{0}/4$ in the both semi-rings whereas the clockwise, for example,  persistent current is directed from left to right in the upper semi-ring but from right to left in the lower semi-ring \cite{Letter07}. 

The challenge to the force balance, $RI_{p} \neq 0$ at $d\Phi /dt = 0$, in the case of superconductor ring is observed only in a narrow critical region $|T - T_{c}| < \delta T_{c} \ll T_{c}$ near the transition into the normal state $T \approx  T_{c}$ where thermal fluctuation or external noise switch ring segments between superconducting $n_{s} > 0$, $R_{s} = 0$ and normal $n_{s} = 0$, $R_{s} > 0$ states. At lower temperature $T <  T_{c} - \delta T_{c}$, in superconducting state when $n_{s} > 0$ all time, $I_{p} \neq 0$ but $R = 0$ whereas at higher temperature $T >  T_{c} + \delta T_{c}$ when $n_{s} = 0$ all time $R > 0$ but $I_{p} = 0$. The circular current should vanish 
$$I(t)  = I_{p}\exp{-\frac{t}{\tau _{LR}}} \eqno{(7)}$$
and the potential difference $V_{s}(t) = R_{s}I(t)$ should be observed during the relaxation time $\tau _{RL} = L/R_{s}$ when any ring segment $l_{s}$ will be switched in the normal state $n_{s} = 0$ with a resistance $R_{s}$. The pair velocity $v$ in the segment $l_{oth} = l - l_{s}$ remaining in superconducting state decreases in accordance with the Newton's law $mdv/dt = qE$ because of the force $qE$ of the potential electric field $E = \nabla V_{s}(t)$. The velocity $v$ should return to the initial value $v \propto n -\Phi /\Phi _{0}$ because of the quantization (6) after the return of the ring segment $l_{s}$ in the superconducting state. This force-free velocity change can explain the absence of the force balance revealed at the $RI_{p}\neq 0$ and $V_{dc}(\Phi /\Phi _{0})$ observations \cite{Letter07}. The pair angular momentum changes from $\oint_{l} dl p = \oint_{l} dl (mv +qA) = q\Phi $ in the state with the unclosed wave function when $v = 0$ to $\oint_{l} dl p = n2\pi \hbar $ (3) in the state with the closed wave function. This change equals $n2\pi \hbar - q\Phi = 2\pi \hbar(n - \Phi /\Phi _{0})$ at each closing of the wave function and  
$$\oint_{l} dl F_{q} = 2\pi \hbar(\overline{n} - \frac{\Phi }{\Phi _{0}})\omega _{sw} \eqno{(8)}$$ 
in a time unity at switching between superconducting states with different connectivity of the wave function with a frequency $\omega _{sw} < 1/\tau _{RL}$. The momentum change in a time unity $F_{q}$ because of the Bohr quantization (3) was called quantum force in \cite{QF2001}. 

The quantum force (8) is not equal zero because of the strong discreteness of permitted state spectrum of superconducting ring $N_{s}(\hbar^{2} /r^{2}m)  \gg k_{B}T$ since the pair number $N_{s}$ is very great in real superconductor ring  \cite{QF2001}. The average value $\overline{n}$ of the quantum number $n$ is close to the integer number $n$ corresponding to minimum energy, i.e. the minimum of the $(n - \Phi /\Phi _{0})^{2}$ value  \cite{QF2001}. Therefore sign and value of the quantum force $F_{q}(\Phi /\Phi _{0})$, as well as $I_{p}(\Phi /\Phi _{0})$ are periodical function of the magnetic flux $\Phi$ with period equal the flux quantum $\Phi _{0}$. The circular quantum force $F_{q}$ takes the place of the circular Faraday's electric field $E = - dA/dt$ and reestablishes the force balance. The voltage oscillations $V_{dc}(\Phi /\Phi _{0}) \propto I_{p}(\Phi /\Phi _{0})$ observed in \cite{Letter07} was predicted \cite{LTPh1998} as consequence of a reiterate switching of the same ring segment into the normal state with a frequency $\omega _{sw}$. According to (7) the voltage average in time should be equal $V_{dc} \approx  L\omega _{sw}\overline{I_{p}}$ at a low frequency $\omega _{sw} \ll  1/\tau _{RL}$ and $V_{dc} \approx  R_{s} \overline{I_{p}}$ at a high frequency $\omega _{sw} \gg  1/\tau _{RL}$.

The quantum force describes only but can not explain the force-free momentum transfer $2\pi \hbar(n - \Phi /\Phi _{0})$ observed at closing of superconducting state. The force-free momentum transfer \cite{FFMT1997} was revealed in the Aharonov-Bohm effect considered for the case of the two-slit interference experiment as far back as 1959 \cite{Bohm1959}. Aharonov and Bohm \cite{Bohm1959} have attracted considerable attention to the problem of the quantum effects of the electromagnetic fluxes \cite{RMP85AhB}. Therefore many authors consider such effects, including the Little-Parks effect \cite{RMP85AhB} and the persistent current \cite{RMP85AhB,PRB07AhB}, as the Ahronov-Bohm effect. The consideration of the Ahronov-Bohm effect in the two-slit interference experiment and in superconducting ring is most suitable way in order to understand the fundamental difference between the quantum force in superconductor \cite{QF2001} and the Bohm quantum potential \cite{Bohm1952}.
 
\subsection{The Aharonov - Bohm effect}
The Aharonov-Bohm effects result directly from the universally recognized interpretation (1) of the wave function phase $\varphi $. But it is no coincidence that this effect was described \cite{Bohm1959} by Bohm who have introduced the quantum potential \cite{Bohm1952}. The Ahronov-Bohm effect demonstrates a non-local force-free momentum transfer \cite{FFMT1997}, which can be described \cite{RMP85AhB,Bohm1982} with help of the non-local quantum potential introduced by Bohm. 

\subsubsection{Non-local force-free momentum transfer in the two-slit interference experiment}
Y. Aharonov and D. Bohm considered two versions of the effect of the electromagnetic potential on the two-slit interference pattern: a magnetic and an electric \cite{Bohm1959}. In the magnetic version a magnetic solenoid situated between the two slits (Fig. 1) produces within it a magnetic flux $\Phi $. The probability 
$$P(y) = |\Psi _{Sh1}|^{2} + |\Psi _{Sh2}|^{2} + 2 |\Psi _{Sh1}| |\Psi _{Sh2}| cos(\varphi _{1} - \varphi _{2}) \eqno{(9)}$$
to finding the particle with a charge $q$ (for example electron with $q = e$) at a point $y$ of the detector screen should depend on the magnetic flux $\Phi $ since the phase difference 
$$\varphi _{1} - \varphi _{2} = \oint_{l}dl \nabla \varphi = \oint_{l}dl \frac{mv + qA}{\hbar } = \Delta \varphi _{0} + 2\pi \frac{\Phi }{\Phi _{0}} \eqno{(10)}$$
(see Fig.1) according to (1). The shift in phase $2\pi \Phi /\Phi _{0}$ observed in the shift of the interference pattern $P(y)$ (9) may be interpreted as momentum transfer since $p = \hbar \nabla \varphi $. This transfer is non-local since the particle never enters the region containing magnetic field and it is force-free since magnetic flux is constant in time $d\Phi /dt = 0$. 

\begin{figure}[b]
\includegraphics{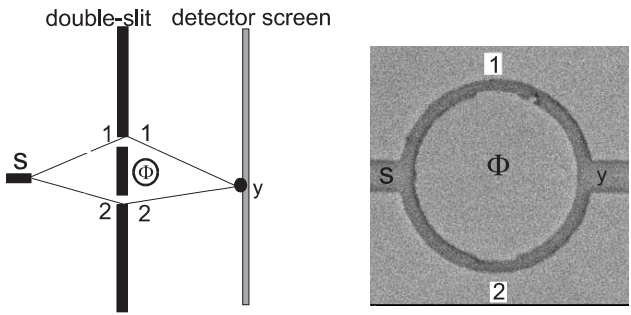}
\caption{\label{fig:epsart} The Ahronov-Bohm effects in the two-slit interference experiment (on the left) and in superconductor ring (on the right) are consequence of the magnetic flux $\Phi $ influence on the phase difference $\varphi _{1} - \varphi _{2} = \oint_{l}dl \nabla \varphi $. In the first case the difference $\varphi _{1} - \varphi _{2} $ of the phase changes between $S$ and $y$ points along upper (1) way $\varphi _{1}$ and lower (2) way $\varphi _{2}$ should not be divisible by $2\pi $ ($\varphi _{1} - \varphi _{2} \neq n2\pi $) since the Schrodinger wave function collapses at particle observation in the two-slit interference experiment. In contrast to the interference experiment the Ahronov-Bohm effects in superconductor ring (for instance the persistent current) are observed since the GL wave function can not collapse and therefore $\varphi _{1} - \varphi _{2} = n2\pi $. }
\end{figure}

\subsubsection{The GL wave function can not collapse }
The phase difference (10) can have any value since the Schrodinger wave function collapses in the interference experiment and therefore $\Psi _{Sh1}$ and $\Psi _{Sh2}$ must not be single-valued at the same point $y$, $\Psi _{Sh1} \neq \Psi _{Sh2}$ and therefore $\varphi _{1} - \varphi _{2} \neq n2\pi $. The Ahronov-Bohm effect in the two-slit interference experiment and the persistent current are observed because of the same relation (1). We can describe superconducting state in two halves of the ring, Fig.1, with two GL wave functions $\Psi _{GL1}$ and $\Psi _{GL2}$ as well as in the interference experiment. But we should write $\Psi _{GL1} = \Psi _{GL2}$ both for the $S$ and $y$ points since the GL wave function can not collapse and therefore it must be single-valued at any point. Just therefore, see the quantization (3), (4), (6), the persistent current and other macroscopic quantum phenomena, i.e. the Ahronov-Bohm effects in superconductor ring, are observed. 

The collapse of the Schrodinger wave function at measurement was postulated by von Neumann (von Neumann's projection postulate) \cite{Neumann}. But physics is empirical science. Therefore it is more important that the collapse of wave function can be observed, for example, in the two-slit interference experiment. It is observed, for example in \cite{Tonomura}, that electron passes through two-slits as a de Broglie wave with wavelength $\lambda _{deB} = 2\pi \hbar /p$ (otherwise the interference pattern (9) could not be observed) which collapses in a point $y$ on the detector screen at electron observation. The collapse of the Schrodinger wave function is observed in many phenomena. But no macroscopic quantum phenomena gives evidence of any wave function collapse at observation. Moreover, the collapse of the wave function representing no a probability $|\Psi _{Sh}|^{2} = P$ but the real density  $|\Psi _{GL}|^{2} = n_{s}$ seems impossible in essence. The density of superconducting pairs $n_{s}$ can not change because of our look.

\subsubsection{Non-local force-free momentum transfer in superconductor ring}
The density $|\Psi _{GL}|^{2} = n_{s}$ can be change because of a real physical cause, for example a heating. We can heat, for example, a small segment $l_{s} \ll  l$ of the superconductor ring shown on Fig.1 up to $T > T_{c}$. There is fundamental difference between our possibility in atom and in superconductor ring \cite{FFP2006}. Both the persistent current in superconductor ring and the stationary orbits in atom are observed because of the Bohr's quantization (3). But in contrast to atom we can enough easily switch superconductor ring between states with different connectivity of the GL wave function. After a heating of only small segment $l_{s} \ll  l$, for example near the point 2 on Fig.1, up to $T > T_{c}$ the circular current $I(t) = sqn_{s}v(t)$ and the pair velocity $v(t)$ in all points of other segment $l - l_{s}$, including the point 1, should decrease down to zero in accordance with the relaxation law (7). In this case the $v(t)$ value changes because of the real force $qE = qR_{s}I(t)/(l - l_{s})$ in agreement with the Newton's law. But the reverse change of the velocity from $v = 0$ to $v = (\hbar /rm)(n - \Phi /\Phi _{0})$ is force-free. This reverse change must be because of the quantization (6) after a cooling of the $l_{s}$ segment down to $T < T_{c}$. This force-free momentum transfer $mv = (\hbar /r)(n - \Phi /\Phi _{0})$ is non-local since an action in the point 2 affects on the velocity $v$ and current $I_{p}$ in the distant point 1 without any field.

\section{Non-universality}
The Bohm's quantum potential and the quantum force in superconductor can describe the non-local force-free momentum transfer revealed with the Aronov-Bohm effects. But the quantum force in superconductor differs in essence from the Bohm's quantum potential because of the different essence of the GL and Schrodinger wave functions \cite{QTRF2007}.  Both wave functions describe quantum phenomena but they can not be universally interpreted. The consideration of macroscopic quantum phenomena together with other one should convince anybody that quantum phenomena can not have an universally interpretation and quantum principles can not be applied universally. For example, macroscopic quantum phenomena refute the universality of the correspondence principle. This principle formulated by Niels Bohr states that the predictions of quantum mechanics reduce to those of classical physics when a system moves to higher energies. How have be able macroscopic quantum phenomena to violate this principle?  

\subsection{Why can macroscopic quantum phenomena be observed?}
The quantum interference of fullerenes and biomolecules with size up to $3 \ nm$ was observed recently \cite{ZeilInEx} and Zeilinger considered \cite{Zeil2004} a possibility of the interference experiment with viruses and nanobacteria. Could we observe the quantum interference of larger systems? 

\subsubsection{We have not the ghost of a chance to observe the quantum interference of a system larger $10^{-6} \ m$ }
In order to observe the two-slit interference pattern of particle with size $a$ its period $\lambda _{deB}L/d$ should be larger $a$. A slit width and a distance between slits $d$ can not be smaller the particle size $a$. Therefore the distance $L$ between the double-slit screen and the detector screen, Fig.1, should be larger than $L = a^{2}/\lambda _{deB}$. Particles pass this distance during a time $t = L/v$ at a velocity $v$. Therefore the interference experiment should take a time $t_{exp} > a^{2}/\lambda _{deB}v = a^{5}g/2\pi \hbar $ since the de Broglie wavelength $\lambda _{deB} = 2\pi \hbar /mv$ and the particle mass $m \approx  ga^{3}$. The value $g/2\pi \hbar \approx 1.5 \ 10^{36} \ c/m^{5}$ at the typical mass density $g \approx  10^{3} \ kg/m^{3}$ of all matter including viruses and bacteria. Thus, the interference of particle with $a < 4 \ 10^{-8} \ m = 40 \ nm$ can be observed at $t_{exp} > 1 \ c$ and the interference experiment at the particle size $a = 10^{-6} \ m = 1 \ \mu m$ should take $t_{exp} > 1$ year, $a = 10 \ \mu m$ - $t_{exp} > 3000$ years; $a = 0.1 \  mm$ - $t_{exp} > 3 \ 10^{8}$ years and so on. 

\subsubsection{Conditions needed for observation of the Bohr quantization }
In order to observe the Bohr's quantization of a particle with a mass $m$ in a ring with radius $r$ the energy difference between permitted states (6)
$$\Delta E_{n+1,n} = \frac{mv_{n+1}^{2}}{2}  - \frac{mv_{n}^{2}}{2} \approx \frac{\hbar ^{2}}{2mr^{2}} \eqno{(11)}$$ 
must exceed the energy of thermal fluctuation $\Delta E_{n+1,n} > k_{B}T$. The discrete spectrum with $\Delta E_{n+1,n} \approx 5 \ 10^{-27} \ J$ of single electron $m = 9 \ 10^{-31} \ kg$ in a ring with radius $r = 10^{-6} \ m = 1 \ \mu m$ can be observed only at very low temperature $T < \Delta E_{n+1,n}/k_{B} \approx  0.0004 \ K$. Therefore first attempts to observe the persistent current in normal metal and semiconductor rings with $r \approx  1 \ \mu m$ were made only twenty years ago \cite{normal,semi}. Recently the persistent current was observed in semiconductor rings with $r \approx  10 \ nm$ at the temperature up to $T = 4.2 \ K$ \cite{semi10nm}. But we have not the ghost of a chance to observe the discrete spectrum in a ring of even virus with $a = 100 \ nm$. Since the ring radius $r$ should exceed the virus size $a$ the temperature of the experiment should be unachievable low $T_{exp} < \Delta E_{n+1,n}/k_{B} < \hbar ^{2}/k_{B}2ga^{5} = 3 \ 10^{-49}/a^{5} \ K/m^{5} = 3 \ 10^{-14} \ K$ at $a = 10^{-7} \ m$. 

\subsubsection{Landau's postulate}
Thus, the estimations show that the correspondence principle should be valid. But in spite of these estimations macroscopic quantum phenomena, superconductivity and superfluidity, are observed. Superconductivity was discovered in 1911 by Heike Kamerlingh Onnes and superfluidity of $^{4}He$ liquid was discovered after 26 years by Pyotr Kapitsa. First successful description was proposed for superfluidity in 1941 by Lev Landau \cite{Landau41}. Landau accentuated in \cite{Landau41} that the Bose-Einstein condensation, described in 1925, can not explain the superfluidity phenomena. In order to describe this macroscopic quantum phenomena he postulated an spectrum of excitations with energy gap. There is important to accentuate that no excitation of individual $^{4}He$ atoms is in this spectrum. Landau postulated virtually that atoms in superfluid $^{4}He$ can not have individual velocity and superfluid condensate moves as one big particle. He wrote in \cite{Landau41} the relation (2) for superconducting current using virtually this postulate for electrons in superconductor. There is important to note that this Landau's postulate should be applied also for superconducting pairs. The energy gap of the Bardeen-Cooper-Schrieffer theory \cite{BCS} concerns the energy spectrum of electrons but not pairs. 

The Landau's postulate about superconducting condensate as a one big particle is implied in the Ginzburg-Landau theory \cite{GL1950}. The wave functions of one free electron $\Psi _{Sh} = |\Psi _{Sh}|\exp{i\varphi }$, where $\nabla \varphi = d\varphi /dl = 2\pi n/l = n/r$, and superconducting condensate $\Psi _{GL} = |\Psi _{GL}|\exp{i\varphi }$,  where $\nabla \varphi = n/r$, in an one-dimensional ring with the radius $r$ look identically. Although $\int_{V}dV |\Psi _{Sh}|^{2} = 1$ and $\int_{V}dV |\Psi _{GL}|^{2} = N_{s} \gg 1$ none of the $N_{s}$ pairs has a quantum number $n$ different from the other $N_{s} - 1$ pairs according to the GL theory. But individual pairs do not vanish totally from superconducting phenomena. The GL theory interpretation of the phase gradient (1) $\hbar \nabla \varphi = \hbar n/r = mv + qA$ as the momentum of single pair ($q = 2e$) but not all $N_{s}$ pairs ($q = 2eN_{s}$) results to the period $\Phi _{0} = 2\pi \hbar /2e$ of the quantum oscillations $I_{p}(\Phi /\Phi _{0})$, $R(\Phi /\Phi _{0})$, $V_{dc}(\Phi /\Phi _{0})$ in magnetic field $\oint_{l}dlA = 2\pi rA = \Phi $ observed in all experiments. The two particles with mass $M = N_{s}m$ of all $N_{s}$ pairs and $m$ of single pair implied in the GL theory allow to explain why macroscopic quantum phenomena can be observed. The difference between the permitted velocity values $v_{n+1} - v_{n} = \hbar /mr$ depends on the mass $m$ of single pair (6) according to (1). Whereas the energy difference between permitted states $E_{n} = Mv_{n}^{2}2/2$ depends also on the mass $M$ of all $N_{s}$ pairs since the quantum number $n$ of an individual pair can not change according to the Landau's postulate. Therefore the discreteness of superconductor ring permitted state spectrum 
$$\Delta E_{n+1,n} = \frac{Mv_{n+1}^{2}}{2}  - \frac{Mv_{n}^{2}}{2} \approx \frac{M}{m}\frac{\hbar ^{2}}{2mr^{2}} = N_{s}\frac{\hbar ^{2}}{2mr^{2}} \eqno{(12)}$$  
increases with the increase of all three sizes of the ring $\Delta E_{n+1,n} \approx  n_{s}s2\pi r(\hbar ^{2}/2mr^{2}) \propto  (s/r)$. Therefore the persistent current is observed in superconductor with how any large sizes. 

\subsection{Quantum description of phenomena should not be universal}
Our comparison of the quantum phenomena described by the Schrodinger wave function and GL wave function has revealed non-universality of quantum description. Some quantum postulates, the correspondence principle, the wave function collapse can not be applied to the description of macroscopic quantum phenomena and the additional Landau postulate is needed for this description. The non-universality may be general rule of whole quantum description. Numerous interpretations prevent to say that the quantum formalism can describe an unique reality and after the violation \cite{BellExp} of the Bell's inequality we can not be sure that quantum phenomena can reflect an unique reality. Anybody who would state about a quantum reality should say what is this reality. Could we be sure that it is a reality of many worlds by Everett \cite{Everett} and others \cite{DeWitt,Deutsch}, non-local reality by Devid Bohm or any others? We can not demand the universality of quantum principles till we can not be sure that quantum formalism describes an unique reality since the universality demand presupposes an unique subject of description.

Many experts interpret the quantum formalism as a description of our knowledge: Schrodinger, "entanglement of our knowledge" \cite{Schrod35}; Heisenberg, "the probability function describes ... rather our knowledge about a quantum process" \cite{Heisen}; Rudolph Peierls, "the most fundamental statement of quantum mechanics is that the wavefunction represent our knowledge of the system" \cite{MerminQT}; Caslav Brukner and Anton Zeilinger "quantum physics is an elementary theory of information" \cite{BrukZeil}; Christopher Fuchs, "the structure of quantum theory has always concerned information" \cite{FuchsInI}. But this information interpretation also can not be universal since it can not be applied to the macroscopic quantum phenomena. Therefore we must admit the positivism point of view that the quantum theory describes only phenomena. In fact Bohr upheld just this point of view. But he did not follow to the corollaries of this positivism position. If anybody has admitted that quantum mechanics can describe only phenomena he should not demand universality of quantum principles. But Bohr upheld in his debate with Einstein \cite{Bohr35,Bohr49,Bohr} and besides this the universality of such principles as complementarity, uncertainty relation and others.

\subsubsection{One more challenge to the universality of the Heisenberg uncertainty relation}
The Bell's experiments \cite{BellExp} have revealed violation one of the EPR assumptions \cite{EPR35}, on the existence of the elements of the physical reality or locality. But it is not correct to conclude that these experiments have proved the universality of the uncertainty principle. Andrei Khrennikov states in \cite{Andrei03} that "(at least in theoretical models) ... the Heisenberg uncertainty principle can be violated". This statement can be corroborate by a very simple Gedankenexperiment which can be made real. The resent experiment \cite{Zeilin02} has corroborated the Heisenberg uncertainty relation for the coordinate $x$ and the velocity $v_{x}$ of fullerene molecules across their motion along $z$. But this result can not mean that this relation will be valid also for the coordinate $z$ and the velocity $v_{z}$ along the motion direction. Moreover it seems obvious that the coordinate $z$ and the velocity $v_{z}$ can be measured simultaneously with any exactness. In order to find the velocity $v_{z} = (z_{2} - z_{1})/(t_{2} - t_{1})$ we should measure two time moments $t_{1}$, $t_{2}$ and two coordinates $z_{2}$, $z_{1}$. The velocity $v_{z}$ can not change between these measurements at the free flying because of the law of momentum $p_{z} = mv_{z}$ conservation. Therefore the differences $z_{2} - z_{1}$ and $t_{2} - t_{1}$ can be made how any large and much larger any measurement inaccuracy of the time moments $\Delta t = \Delta t_{1} + \Delta t_{2}$ and the coordinates $\Delta z = \Delta z_{1} + \Delta z_{2}$. We can take $t_{1} = 0$, $t_{2} = t$ and $x_{1} = 0$, $x_{2} = x$. The velocity value $v_{z} = z/t$ can be measured with the uncertainty $\Delta v_{z} \approx v_{z}(\Delta z/z + \Delta t/t)$ at $x \gg  \Delta x$, $t \gg  \Delta t$ and the product of the velocity $\Delta v_{z}$ and coordinate $\Delta z$ uncertainties 
$$\Delta z  \Delta v_{z} \approx \Delta z  v_{z}( \frac{\Delta z}{z} + \frac{\Delta t}{t})  \eqno{(13)}$$  
can be made how any small since the $z$ and $t$ values can be made how any large. 

Thus, the Heisenberg uncertainty relation $\Delta z \Delta v_{z} > \hbar /2m$ should be violated if we use the method of the velocity measurement learned in the primary school. This method may be used in an actual experiment and in order to verify the uncertainty relation no enormous distance is needed. The value $\hbar /2m \approx  0.3 \ 10^{-10} \ m^{2}/c$ for the fullerene molecules $m \approx  1.4 \ 10^{-24} \ kg$ used in \cite{Zeilin02}. The product $\Delta zv_{z} < 10^{-4} \ m^{2}/c$ at the velocity $v_{z} = 100 \ m/c$ and the measurement inaccuracy of coordinates $\Delta z < 10^{-6} \ m$ in the actual experiment \cite{Zeilin02}. The uncertainty relation should violated according to (13) at $\Delta z/z < 0.3 \ 10^{-6}$ and $\Delta t/t < 0.3 \ 10^{-6}$, i.e. at the values accessible for an actual experiment $z > 3 \ m$ (in \cite{Zeilin02} $z > 2 \ m$) and $\Delta t = 0.3 \ 10^{-6} t = 0.3 \ 10^{-6}z/v_{z} < 10^{-8} \ c$ at $z = 3 \ m$. 

\subsubsection{Could the Bohr's complementarity be considered as an universal principle?}
Bohr postulated the complementarity when he has despaired of quantum world realism. This Bohr's principle has formulated the renunciation of the reality description as the goal of quantum theory. Therefore it can not be universal in principle. Nevertheless most physicists following Bohr believe in its universality. Only some experts discuss  the essence of complementarity \cite{AndreiCo,Plotnits} and some authors \cite{Afshar} challenge this principle. Arkady Plotnitsky \cite{Plotnits} considering "perhaps Bohr's most refined formulation of what he means by the complementary situations of measurement" given in \cite{Bohr49} formulates Bohr's complementarity principle as follows: "in considering complementary conjugate variables in question in quantum mechanics (in contrast to those of classical physics) we deal with two mutually exclusive experimental arrangements, each providing complete (actual) information about the two states of the two quantum objects involved in such measurements". The position $z$ and momentum $p_{z} = mv_{z}$ are referred as example of "complementary conjugate variables" in the "Bohr's most refined formulation" \cite{Plotnits}. But the actual experimental arrangements of the school method considered above are not merely mutually non-exclusive for measurement of position and momentum. They are the same. We should conclude that either the method of the momentum measurement $p_{z} = mv_{z} = mz/t$ learned in the primary school is not correct (but why?!) or the Bohr's complementarity principle is not universal. 

\subsubsection{Logic of universality as consequence of naive realism point of view}
Einstein upheld the principle of realism and he stated that the quantum theory had relinquished precisely what has always been the goal of science: "the complete description of any (individual) real situation (as it supposedly exists irrespective of any act of observation or substantiation)" \cite{Einste49}. In contrast to this Bohr stated that all hope of attaining a unified picture of objective reality must be abandoned. According to his point of view quantum theory would provide predictions concerning the results of measurements but, unlike all previous theories, it was incapable of providing a full account of "how nature did it." Bohr argued that the very desire to seek such a complete account was misguided and naive: "There is no quantum world. There is only an abstract quantum physical description. It is wrong to think that the task of physics is to find out how Nature is" (the citation from \cite{BrukZeil}). 

Most physicists agree rather with Bohr than Einstein since we have no "hope of attaining a unified picture of objective reality" of quantum world up to now. But most physicists remain in fact naive partisan of the realism point of view since it is difficult to renounce really the centuries-old realistic point of view with all that it implies. If anybody has admitted that "there is no quantum world" he should not think that quantum description must be universal. But most physicists are sure that quantum principles must be universal. For example, most authors are sure that any quantum system with two permitted states \cite{Nielsen}, including superconductor ring \cite{Makhlin}, can be used as quantum bit (qubit), principal element of quantum computer. Qubit, in contrast to classical bit, should operate in state superposition. But how could superconductor ring be in state superposition if the GL wave function can not collapse? Superposition of states is inconceivable without the collapse since we can not observe anything in two place at the same time. Therefore Anthony Leggett assuming superposition of superconductor ring states should be to fabricate a new wave function \cite{Leggett}, which can collapse, in additional to the GL wave function, which can not collapse. The new wave function was fabricated for superconductor ring but not for the Schrodinger cat, for example, because of the logic of universality. But this logic should not be applied to quantum formalism which is phenomenological theory.

\section*{Acknowledgement}
This work has been supported by grant "Quantum bit on base of micro- and nano-structures with metal conductivity" of the Program "Technology Basis of New Computing Methods" of ITCS department of RAS, grant 08-02-99042-r-ofi of the  RFBR and grant of the Program "Quantum Nanostructures" of the Presidium of Russian Academy of Sciences.

\end{document}